\documentclass[twocolumn,secnumarabic,amsmath,amssymb,balancelastpage,nofootinbib]{article}

\usepackage[e]{esvect}
\usepackage{color}         
\usepackage{graphics}      
\usepackage{graphicx}      
\usepackage{epsf}          
\usepackage{bm}            

\usepackage{cite}
\usepackage{amssymb}
\usepackage{amsmath, esint}
\usepackage{mathrsfs}
\usepackage{framed}
\usepackage{bigints} 
\usepackage{enumitem}
\usepackage{pifont}
\usepackage[capbesideposition={left,center},facing=yes,capbesidewidth=8cm,capbesidesep=quad]{floatrow} 
\usepackage{setspace} 

\usepackage[none]{hyphenat} 

\usepackage[colorlinks=true]{hyperref}  

\usepackage{scrextend}
\usepackage{graphicx}

\usepackage{afterpage}

\definecolor{darkred}{rgb}{0.6,0,0}
\definecolor{darkgreen}{rgb}{0,0.5,0}
\definecolor{darkblue}{rgb}{0,0,0.6}
\hypersetup{ colorlinks,
linkcolor=darkblue,
filecolor=darkgreen,
urlcolor=darkgreen,
citecolor=darkred }

\begin{document}

\sloppy 

\bibliographystyle{nar}

\newlength{\bibitemsep}\setlength{\bibitemsep}{.2\baselineskip plus .05\baselineskip minus .05\baselineskip}
\newlength{\bibparskip}\setlength{\bibparskip}{0pt}
\let\oldthebibliography\thebibliography
\renewcommand\thebibliography[1]{%
  \oldthebibliography{#1}%
  \setlength{\parskip}{\bibitemsep}%
  \setlength{\itemsep}{\bibparskip}%
}

\title{\vspace*{-35 pt}\huge{Field Velocity and Mass Renormalization\\in Classical Electromagnetism}}
\author{Charles T. Sebens\\Division of the Humanities and Social Sciences\\California Institute of Technology}
\date{arXiv v1 - September 20, 2026}

\twocolumn[
\maketitle     
\vspace*{-20 pt}
\begin{quote}
The intricate procedure of mass renormalization in quantum field theory has a simpler precursor in classical electromagnetism.  There, the renormalized mass of a charged body includes both its bare mass and its electromagnetic mass.  For a body moving at constant low velocity, the electromagnetic mass can be calculated either from the total momentum of the electromagnetic field or from the total energy of the electromagnetic field in the body's rest frame.  For a spherically symmetric charge distribution, these methods appear to disagree by a factor of 4/3.  This 4/3 problem has a long history.  Here, we see that the concept of field velocity can be used to quickly resolve the 4/3 problem.  The two methods for finding the electromagnetic mass agree after noting that the average velocity of energy flow in the electromagnetic field is 4/3 the velocity of the body.  The energy in the electromagnetic field moves faster than the charged body because, in addition to moving with the body, the field energy must make its way from the back of the body (where it is being emitted) to the front (where it is being absorbed).
\end{quote}
\vspace*{20 pt}
]

\tableofcontents

\section{Introduction}

In classical electromagnetism, a charged body at rest is accompanied by an electromagnetic field with its own energy.  By mass-energy equivalence, the energy of this field contributes an electromagnetic mass towards the total or ``renormalized'' mass of the body.  If that charged body is set in motion at constant velocity, its electromagnetic field will carry both energy and momentum.  However, even at low velocities, the momentum of the moving body's field is not equal to the aforementioned electromagnetic mass times the velocity.  If the body has a spherically symmetric charge distribution, the momentum differs by a factor of $4/3$ that has been described as ``infamous'' \cite[pg.\ 755]{jackson1999} and ``notorious'' \cite{griffithsowen1983}.  The long debate over this ``$4/3$ problem'' has largely subsided, with a variety of solutions having been put on the table and many authors agreeing that the $4/3$ factor disappears when we focus our attention on the total mass and momentum of both the electromagnetic field and the body itself (including contributions associated with the non-electromagnetic ``Poincar\'{e} stresses'' holding the body together).  My aim here is not to challenge those accounts regarding the total mass and momentum, but to clarify the relationship between the mass and momentum of the electromagnetic field.

In this article, I first present the $4/3$ problem and show how it can be quickly resolved by introducing a velocity for the electromagnetic field itself and seeing that the electromagnetic field's energy is, on average, moving faster than the charged body.  The electromagnetic field's momentum is equal to its mass times this velocity.  I then ask: How can the cloud of electromagnetic field energy surrounding the body move faster than the body without separating from the body?  That turns out to be possible because energy is being transferred between the field and different parts of the charged body.  In addition to moving with the body, field energy is moving from the back of the body (where the charged body is losing energy to the field) to the front (where the charged body is gaining energy from the field).

Although it is sometimes remarked that one can introduce a velocity for the electromagnetic field, that velocity is rarely used.  Here we see an example of what it can do, illustrating why it might deserve a more prominent place in electromagnetism courses and textbooks.  In the end, we see that the $4/3$ problem does not indicate any inconsistency or incoherence in the classical procedure of mass renormalization.

The analysis of mass renormalization in classical electromagnetism is of particular interest because of the light that it might shed on mass renormalization in quantum field theory (as will be discussed briefly in the conclusion).  For that reason, you may wish to think of the charged body at hand as a classical electron.

\section{A Charged Sphere}

\begin{figure}[htb]
\center{{\large \sc{Electric Field}}}
\center{\includegraphics[width=6 cm]{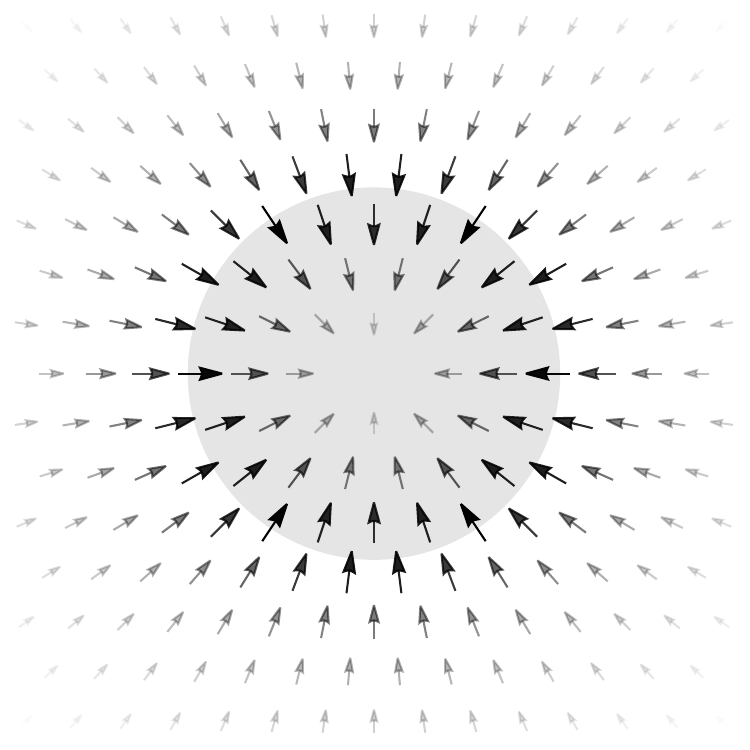}}
\caption{The electric field of a uniformly negatively charged sphere, with field strength indicated by the size and darkness of the arrows.}
  \label{Efig}
\end{figure}

To work with a simple and concrete example, let us take our charged body to be a uniformly charged sphere with radius $R$, total charge $Q$, and thus charge density
\begin{equation}
\rho=\frac{Q}{\frac{4}{3}\pi R^3}
\ .
\label{chargedensitysphere}
\end{equation}
At rest, this sphere comes with an electric field that is strongest at the surface of the sphere (figure \ref{Efig}).  In Gaussian CGS units,
\begin{equation}
\vec{E}(\vec{x})=\begin{cases} \frac{Q}{|\vec{x}|^2}\hat{x} &\mbox{if } |\vec{x}| > R \\
\frac{Q|\vec{x}|}{R^3}\hat{x} & \mbox{if } |\vec{x}| \leq R
\end{cases}
\ ,
\label{Efieldsphere}
\end{equation}
where $\hat{x}$ is the unit vector pointing in the $\vec{x}$ direction.  At any point within the sphere, the charge located farther from the center makes no contribution and the electric field is effectively sourced by the fraction of the total charge that is located at smaller radii, $\frac{Q|\vec{x}|^3}{R^3}$.

By the Lorentz force law,
\begin{align}
\vec{f} = \rho \left(\vec{E} + \frac{1}{c}\vec{v} \times\vec{B} \right)
\ ,
\label{Lorentzforcelaw}
\end{align}
we multiply \eqref{chargedensitysphere} and \eqref{Efieldsphere} to get the density of electromagnetic force on the sphere,
\begin{align}
\vec{f}(\vec{x}) = \frac{3Q^2 |\vec{x}|}{4 \pi R^6}\hat{x}
\ .
\label{selfrepulsion}
\end{align}
This is a force of self-repulsion that must be balanced by non-electromagnetic cohesive forces for the sphere to retain its shape.

Let us assume the standard expressions for the energy density, energy flux density, and momentum density of the electromagnetic field:\footnote{These quantities appear in the standard symmetric energy-momentum tensor for the electromagnetic field.}
\begin{align}
\rho^{\mathcal{E}}&=\frac{E^2}{8\pi}+\frac{B^2}{8\pi}
\label{EMenergydensity}
\\
\vec{S}&=\frac{c}{4\pi}\vec{E}\times\vec{B}
\label{Poyntingvector}
\\
\vec{G}&=\frac{1}{4\pi c}\vec{E}\times\vec{B}
\label{EMmomentumdensity}
\ .
\end{align}
The total electromagnetic energy of the sphere at rest can thus be calculated by integrating $\frac{E^2}{8\pi}$ inside and outside the sphere, yielding
\begin{equation}
\mathcal{E}=\frac{3Q^2}{5 R}
\ .
\label{Eenergysphere}
\end{equation}
Applying mass-energy equivalence and dividing this energy by $c^2$ yields an electromagnetic mass of
\begin{equation}
m_{em}=\frac{3Q^2}{5 R c^2}
\ .
\label{Eenergysphere}
\end{equation}

\begin{figure}[htb]
\center{{\large \sc{Magnetic Field}}}
\center{\includegraphics[width=6 cm]{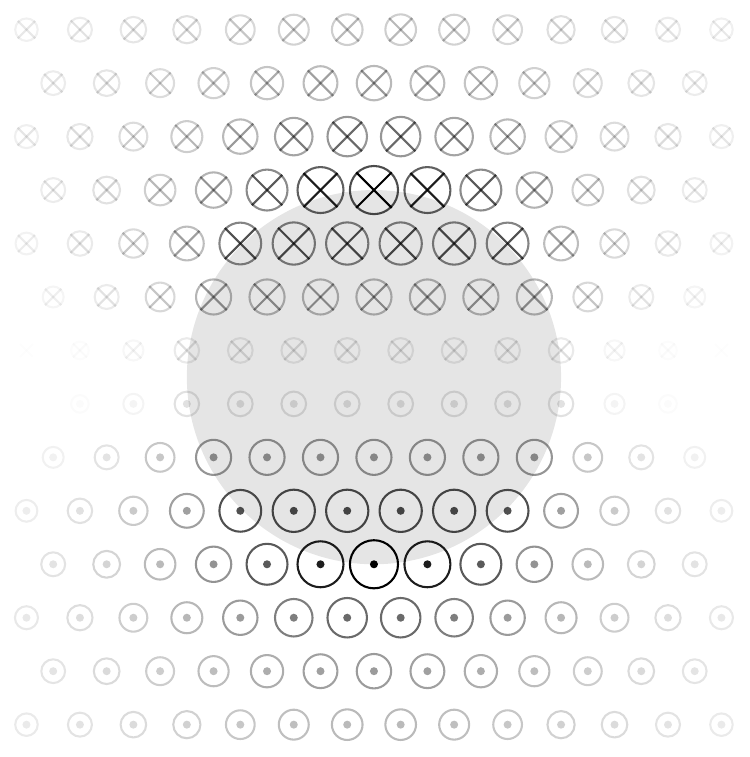}}
\center{\includegraphics[width=4.5 cm]{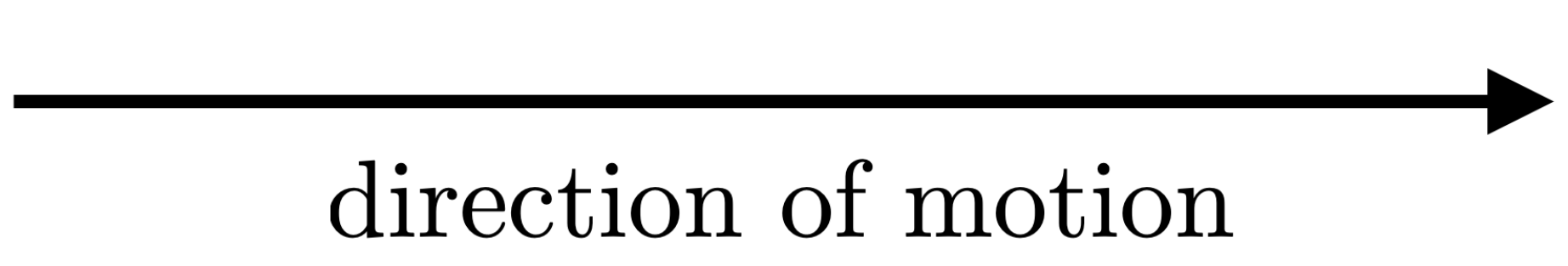}}
\caption{The magnetic field of a negatively charged sphere moving to the right at constant low velocity.  The field lines circle the axis picked out by the sphere's motion, pointing into the page in the top half of the figure and out of the page in the bottom half.}
  \label{Bfig}
\end{figure}

Let us now suppose that the sphere is moving with a constant velocity.  Just as a wire with current flowing through it is surrounded by a magnetic field that circles the wire, our moving sphere is surrounded by a magnetic field that circles its direction of motion (figure \ref{Bfig}).  To explicitly write down that field, we can appeal to the constant-low-velocity ``quasi-electrostatic'' approximation,\footnote{See \cite[examples 14.3 and 15.2]{zangwill2012}.} where we treat each bit of charge as sourcing a Coulomb electric field that moves along with the velocity of that bit of charge \eqref{Efieldsphere} as well as a magnetic field $\vec{B}=\frac{1}{c}\vec{v}\times\vec{E}$, which in our case is
\begin{equation}
\vec{B}(\vec{x})=\begin{cases} \frac{vQ}{c|\vec{x}|^2}\hat{v}\times\hat{x} &\mbox{if } |\vec{x}| > R \\
\frac{vQ|\vec{x}|}{cR^3}\hat{v}\times\hat{x}  & \mbox{if } |\vec{x}| \leq R
\end{cases}
\ .
\label{Bfieldsphere}
\end{equation}
If the charged sphere is moving near the speed of light, the electromagnetic field around it will start to look a bit different (with the electric field lines clumping around the plane perpendicular to the motion and becoming sparser both behind and in front of the sphere).  For our purposes here, we do not need to consider these high-speed effects.

Within the quasi-electrostatic approximation, the total energy of the electromagnetic field can be found by integrating $\frac{E^2}{8\pi}$, as in \eqref{Eenergysphere}, and $\frac{B^2}{8\pi}$ for the magnetic field in \eqref{Bfieldsphere}, yielding
\begin{equation}
\mathcal{E}=\frac{3Q^2}{5 R} + \frac{2Q^2v^2}{5 R c^2}
\ .
\label{totalenergysphere}
\end{equation}
Dividing this quantity by $c^2$ yields a relativistic mass for the electromagnetic field of the moving sphere (the kind of mass that is always proportional to energy).  As we are assuming the velocity to be low, let us set aside terms of order $\frac{v^2}{c^2}$.  Then, the energy density of the field around the moving sphere is the same as the energy density around the stationary sphere, and the electromagnetic relativistic mass is equal to the electromagnetic ``proper'' or ``rest'' mass.  In this article, ``mass'' is used as shorthand for rest mass and relativistic mass is only mentioned occasionally.

\begin{figure}[htb]
\center{{\large \sc{Momentum Density}}}
\center{\includegraphics[width=6 cm]{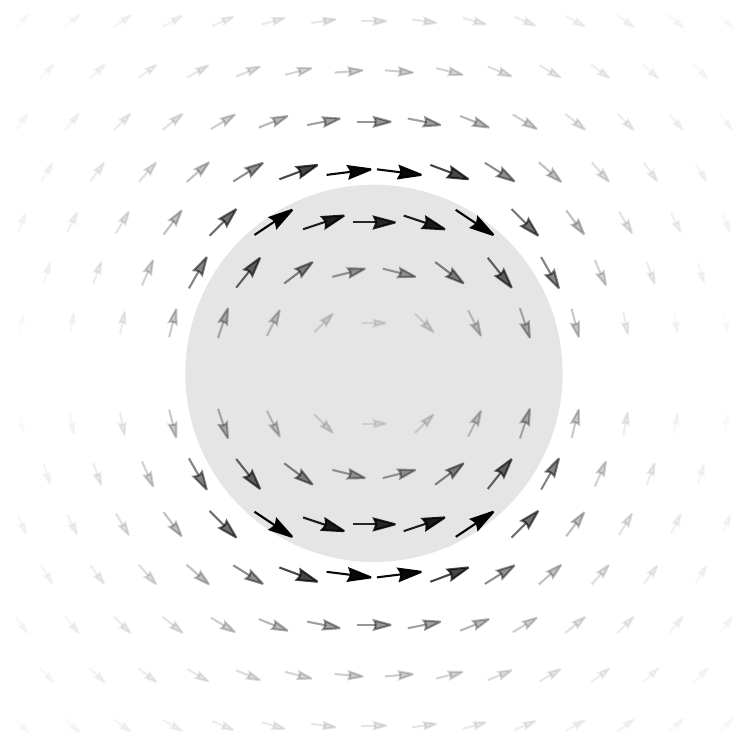}}
\caption{The momentum density of the electromagnetic field around the sphere, found by taking the cross product of the electric field in figure \ref{Efig} and the magnetic field in figure \ref{Bfig}.}
  \label{Gfig}
\end{figure}

The momentum density \eqref{EMmomentumdensity} of the electromagnetic field around the sphere is
\begin{equation}
\vec{G}=
\begin{cases}  \frac{Q^2 v}{4\pi c^2 |\vec{x}|^4} \hat{x} \times (\hat{v}\times\hat{x})&\mbox{if } |\vec{x}| > R \\
\frac{Q^2 v |\vec{x}|^2}{4\pi c^2 R^6} \hat{x} \times (\hat{v}\times\hat{x})
& \mbox{if } |\vec{x}| \leq R
\end{cases}
\ ,
\label{Gsphere}
\end{equation}
where the triple product $\hat{x} \times (\hat{v}\times\hat{x})$ can be alternatively be written as $\hat{v}-(\hat{x} \cdot\hat{v})\hat{x}$.  In spherical coordinates where $\vec{v}$ points in the $z$ direction and $\theta$ is the polar angle off of the $z$ axis (ranging from 0 to $\pi$), the momentum density is
\begin{equation}
\vec{G}=
\begin{cases}  -\frac{Q^2 v}{4\pi c^2 r^4} \sin(\theta)\hat{\theta}&\mbox{if } r > R \\
-\frac{Q^2 v r^2}{4\pi c^2 R^6} \sin(\theta)\hat{\theta}
& \mbox{if } r \leq R
\end{cases}
\ ,
\label{Gsphere2}
\end{equation}
as depicted in figure \ref{Gfig}.  The total momentum can be found by integrating the momentum density:
\begin{equation}
\vec{p}_{em}=\frac{4Q^2}{5 R c^2}\vec{v}=\frac{4}{3}m_{em}\vec{v}
\ .
\label{totalmomentumsphere}
\end{equation}
The infamous factor of $\frac{4}{3}$ has appeared.

\section{The Standard Puzzle}\label{standardpuzzlesection}

With that setup in place, we can now state the $4/3$ problem as it is usually presented \cite{feynman2, schwinger1983, griffithsowen1983, moylan1995, griffiths2012}.  The following two claims may both seem plausible, but are contradictory and thus cannot both be true:
\begin{enumerate}
\item By mass-energy equivalence ($\mathcal{E}=m c^2$), the electromagnetic mass of the charged sphere moving at constant velocity can be found by going to the sphere's rest frame and dividing the total energy in the sphere's electromagnetic field by $c^2$, yielding $m_{em}$ in \eqref{Eenergysphere}.
\item Because (at low velocities) momentum equals mass times velocity, the electromagnetic mass of the charged sphere moving at constant low velocity $\vec{v}$ can be found by calculating the total momentum of the sphere's electromagnetic field \eqref{totalmomentumsphere} and finding the factor preceding $\vec{v}$, yielding $\frac{4}{3}m_{em}$.
\end{enumerate}
The energy-derived electromagnetic mass and the momentum-derived electromagnetic mass differ.\footnote{There is a third way of finding the electromagnetic mass, by looking at the way a body responds to forces, that we will come to at the end of section \ref{resolutionsection}.}  Some explicit or implicit assumption in 1 or 2 must be incorrect.

The most common way to address the problem\footnote{This usual solution is presented and compared to alternatives in, e.g., \cite{griffithsowen1983, campos1986}; \cite[ch.\ 16]{jackson1999}.  The approach traces its heritage back to Poincar\'{e} \cite{poincare1906, janssen2006}.} is to argue that when we carefully consider the energy and momentum of the sphere itself (including the internal energy associated with overcoming the electric self-repulsion in \eqref{selfrepulsion}), the total momentum of the sphere and field system is equal to the total mass of the sphere and field system (derived from the total energy in the rest frame) times the velocity of the system, $\vec{v}$.  The fact that the totals are well-behaved is nice, but does not immediately resolve the puzzle above.  On this approach, the assumption being rejected is presumably that the electromagnetic field’s momentum is equal to electromagnetic mass times velocity (as the electromagnetic contribution to the system's total mass is $m_{em}$ and the electromagnetic contribution to the total momentum is $\frac{4}{3}m_{em}\vec{v}$).  Because the total mass of the system is the sum of the electromagnetic mass and the mass of the sphere itself,
\begin{equation}
m=m_{em}+m_{s}
\ ,
\end{equation}
and total momentum is the sum of the field's momentum and sphere's: if the total momentum of the system is $m\vec{v}$ and the field's momentum is not $m_{em}\vec{v}$, then the sphere's momentum must also fail to be equal to the sphere's mass times velocity, $m_{s}\vec{v}$.

\section{Field Velocity and a Revised Puzzle}

The total momentum in the electromagnetic field of the moving sphere is not what you would have expected from multiplying the rest-energy-derived electromagnetic mass by the sphere's velocity $\vec{v}$.  There are three possible responses.  First, you might infer that the electromagnetic mass is not the rest-energy-derived electromagnetic mass $m_{em}$, as in the second part of the puzzle in the previous section.  Second, you might infer that the electromagnetic momentum is not electromagnetic mass times velocity, as in the resolution presented at the end of the previous section.  Third, it could be that the velocity of the electromagnetic field is not the same as the sphere's velocity $\vec{v}$.  To see if this is the case, let us examine the velocity of the electromagnetic field itself.

Dividing the electromagnetic field's energy flux density \eqref{Poyntingvector} by its energy density \eqref{EMenergydensity} gives a general expression for the velocity of energy flow at each point in space,
\begin{equation}
\vec{v}_f=\frac{\frac{c}{4 \pi} \vec{E} \times \vec{B}}{\frac{E^2}{8\pi}+\frac{B^2}{8\pi}}=2 c \frac{\vec{E} \times \vec{B}}{E^2+B^2}
\ ,
\label{fieldvelocity}
\end{equation}
that is capped at $c$ when the electric and magnetic fields are perpendicular and equal in strength.\footnote{This velocity appears in \cite{poincare1900};\cite{geppert1965, arora1967}; \cite[sec.\ 12.6.2]{holland1993}; \cite[sec.\ 15.2.1]{bornwolf1999}; \cite[box 8.3]{lange2002}; \cite{forcesonfields2018, howelectronsspin, gravitationalfield2022}.}

\begin{figure}[htb]
\center{{\large \sc{Field Velocity}}}
\center{\includegraphics[width=6 cm]{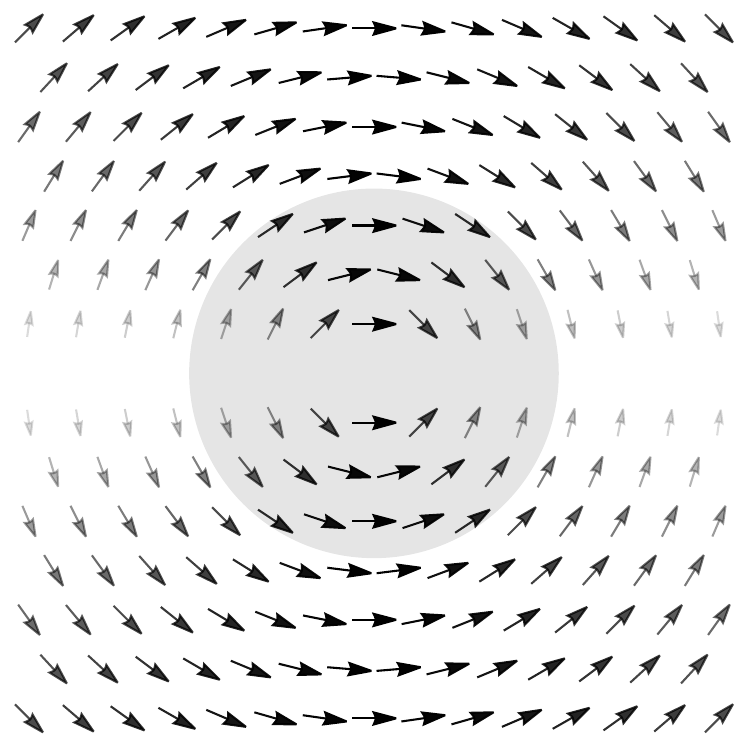}}
\caption{The electromagnetic field velocity around the sphere.}
  \label{Vfig}
\end{figure}

For the charged sphere moving at constant low velocity, the field velocity is
\begin{align}
\vec{v}_f&=2 v \ \hat{x} \times (\hat{v}\times\hat{x})
\nonumber
\\
&= 2 v (\hat{v}-(\hat{x} \cdot\hat{v})\hat{x})
\ ,
\end{align}
both inside and outside the sphere (figure \ref{Vfig}).  (The $\frac{B^2}{8\pi}$ term in the denominator of \eqref{fieldvelocity} is of order $\frac{v^2}{c^2}$ and has been dropped.)  In spherical coordinates, the field velocity is
\begin{align}
\vec{v}_f&=- 2 v \sin(\theta)\hat{\theta}
\ .
\end{align}
We can find the average velocity by inserting an energy weighting and integrating over all space
\begin{align}
\langle \vec{v}_f \rangle&=\frac{1}{\mathcal{E}}\int d^3\vec{x}\  \rho^{\mathcal{E}} \vec{v}=\frac{4}{3}\vec{v}
\ .
\end{align}
Note that the integrand is the energy flux density (which is proportional to the momentum density).  In this case and in general, the average velocity $\langle \vec{v}_f \rangle$ can be multiplied by the field's relativistic mass, $\frac{\mathcal{E}}{c^2}$, to get the total momentum in the electromagnetic field, $\int d^3\vec{x}\ \vec{G}=\int d^3\vec{x}\ \frac{1}{c^2}\rho^{\mathcal{E}} \vec{v}$,
\begin{equation}
\vec{p}_{em}=\frac{\mathcal{E}}{c^2}\langle \vec{v}_f \rangle
\ .
\label{totalmomentum}
\end{equation}
In the low-velocity approximation used here, the rest mass and relativistic mass are the same, as was noted under \eqref{totalenergysphere}, and the field's total momentum is equal to its mass $m_{em}$ times its velocity $\frac{4}{3}\vec{v}$. Beyond that approximation, \eqref{totalmomentum} extends the idea that momentum equals mass times velocity to the context of a relativistic classical field theory.

This analysis of the electromagnetic field's velocity resolves the puzzle from the previous section, rejecting the assumption in part 2 that the relevant velocity for determining the field's mass from its momentum is the sphere's velocity.  However, this resolution raises a new puzzle: If the electromagnetic field's energy is (on average) moving faster than the charged sphere, why does the cloud of field energy remain centered on the sphere as the sphere moves?

\section{The Resolution}\label{resolutionsection}

\begin{figure}[htb]
\center{\includegraphics[width=8 cm]{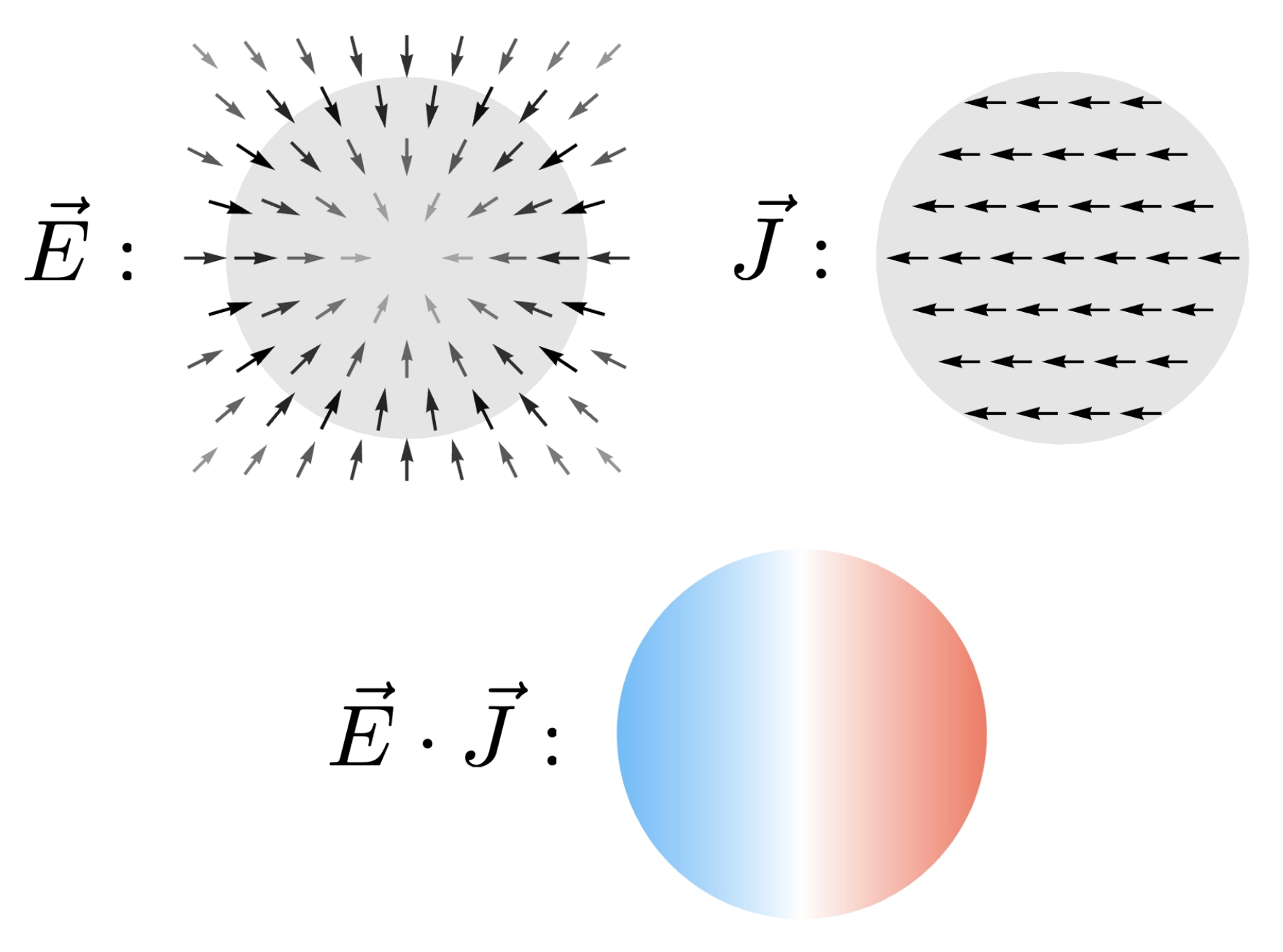}}
\caption{The electric field, current density, and dot product $\vec{E}\cdot\vec{J}$ (describing the rate at which energy is transferred from the electromagnetic field to charged matter per unit volume).  The red indicates a flow of energy from the field into the sphere and the blue indicates a flow out from the sphere.}
  \label{EJfig}
\end{figure}

\begin{figure}[htb]
\center{{\large \sc{Field Velocity Minus $\vec{v}$}}}
\center{\includegraphics[width=6 cm]{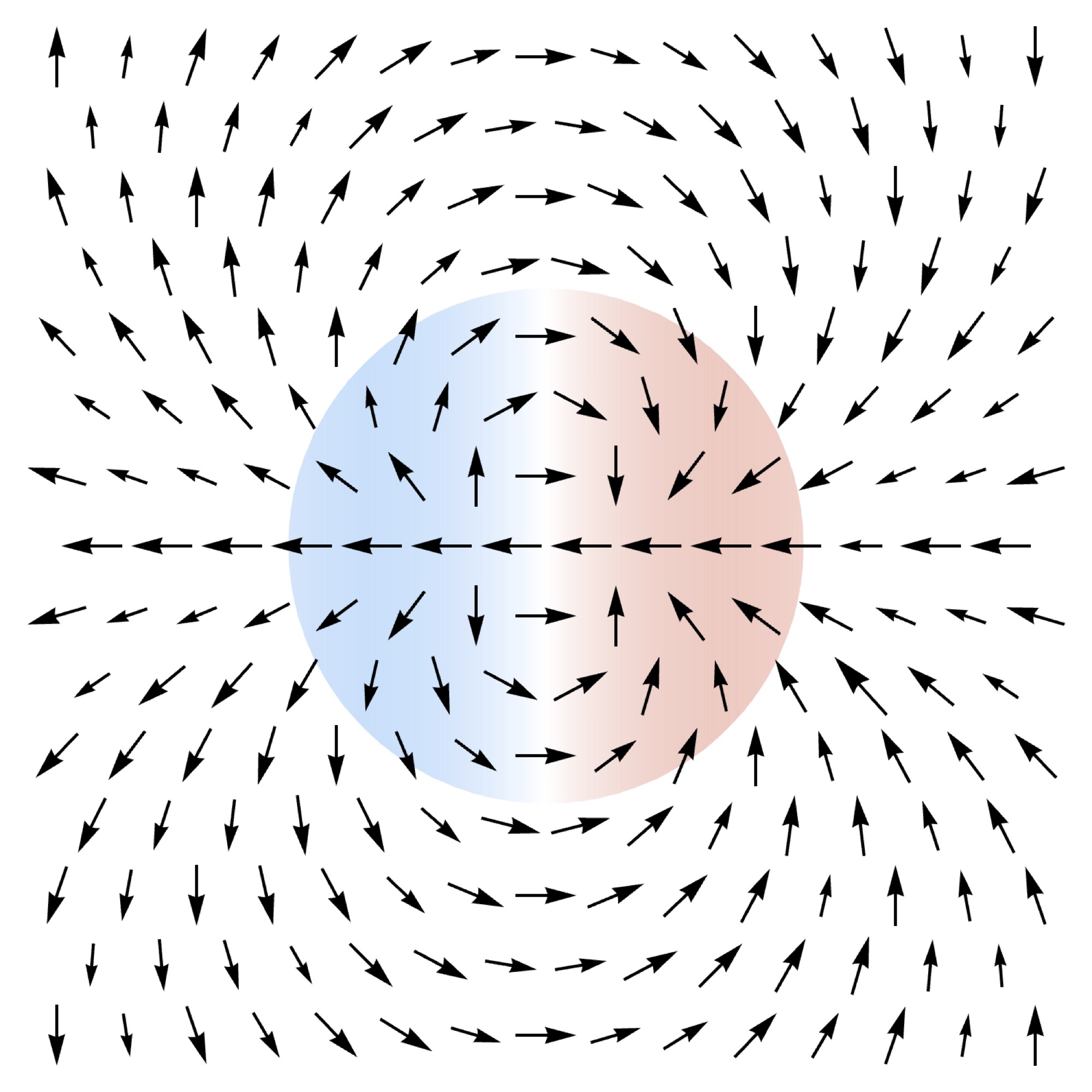}}
\caption{The velocity of the electromagnetic field of the moving sphere, at each point subtracting out the velocity $\vec{v}$ at which the sphere is moving forward (to show the circulation of energy through the sphere).  Note that this is not the field velocity in the reference frame of an observer moving with the sphere, which would be everywhere zero.}
  \label{V2fig}
\end{figure}

\begin{figure}[htb]
\center{{\large \sc{Momentum Density\\Minus $\left(\rho^{\mathcal{E}}/c^2\right)\vec{v}$}}}
\center{\includegraphics[width=6 cm]{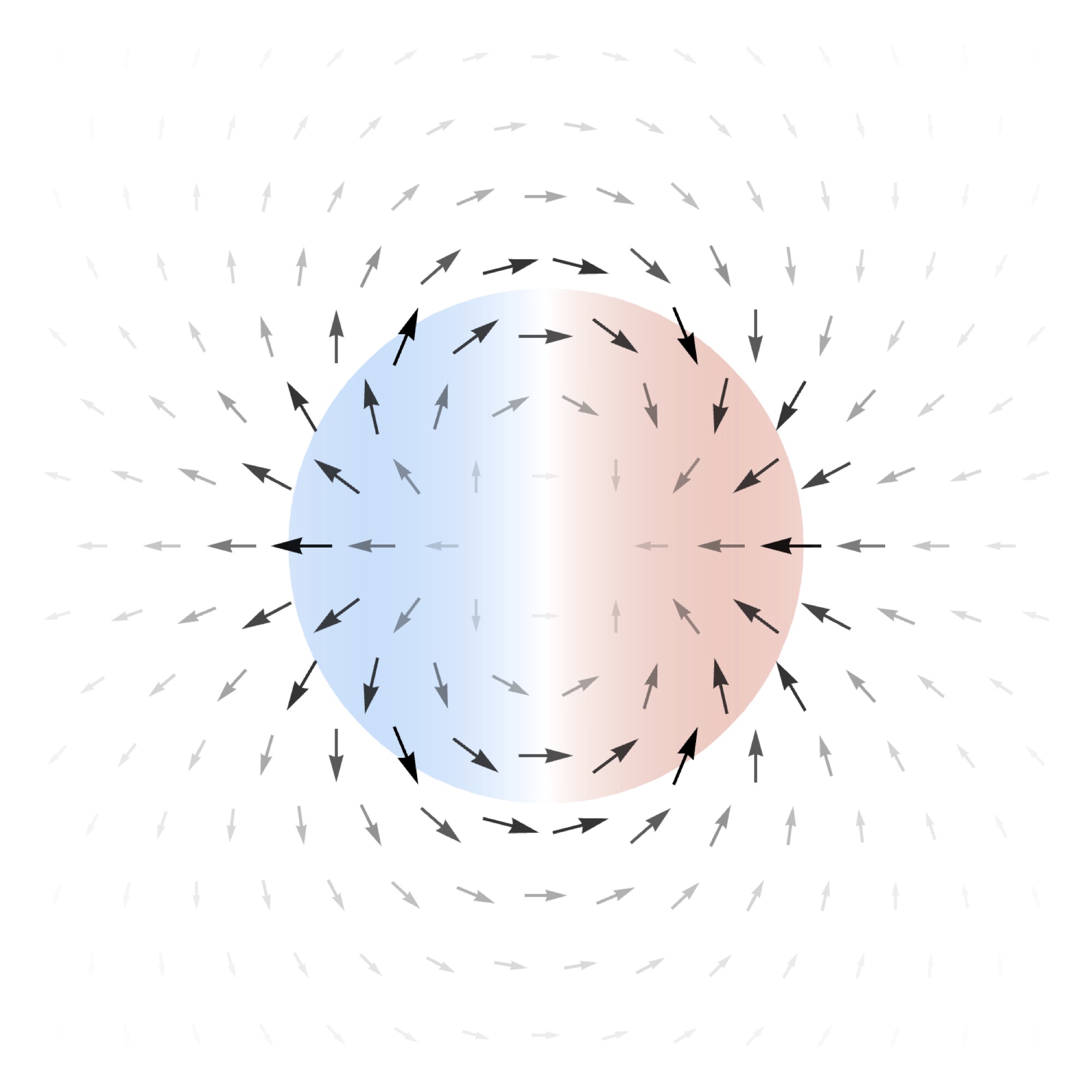}}
\caption{The momentum density of the electromagnetic field, subtracting out the momentum density that you would have if the field mass moved along at the same velocity as the sphere: $\frac{\rho^{\mathcal{E}}}{c^2}\vec{v}$.}
  \label{G2fig}
\end{figure}

The revised puzzle can be quickly solved by analyzing the equation for the local conservation of electromagnetic energy,
\begin{align}
\frac{\partial \rho^{\mathcal{E}}}{\partial t}&= -\vec{\nabla}\cdot\vec{S}-\vec{E}\cdot\vec{J}
\nonumber
\\
&= -\vec{\nabla}\cdot(\rho^{\mathcal{E}} \vec{v}_f)-\vec{E}\cdot\vec{J}
\ ,
\label{energyconservation}
\end{align}
where $\vec{J}$ is the current density and $\vec{E}\cdot\vec{J}$ is the rate at which energy is transferred from the electromagnetic field to matter per unit volume.  For our moving charged sphere, $\vec{J}$ is equal to $\rho\vec{v}$, with $\rho$ given by \eqref{chargedensitysphere}.  The $-\vec{E}\cdot\vec{J}$ source term in \eqref{energyconservation} is non-zero because the forces of electric self-repulsion do work on different parts of the sphere as it moves (figure \ref{EJfig}),
\begin{align}
-\vec{E}\cdot\vec{J}&=-\frac{3 Q^2}{4\pi R^6}\vec{x} \cdot\vec{v}
\ .
\label{EdotJ}
\end{align}
The front half of the charged sphere (where $\vec{x} \cdot\vec{v}$ is positive) absorbs energy from the electromagnetic field and acts a sink  for electromagnetic field energy.  The back half acts as a source, emitting energy into the electromagnetic field.  The front half of the sphere eats energy and the back half spits it out.\footnote{This transfer of energy between field and matter has been emphasized by Morozov \cite{morozov2011} in his analysis of a spherical shell of charge, where he also discusses flow of energy within the sphere from front to back (an idea that he traces back to Becker \cite{becker1933} and appears briefly in \cite[vol.\ 1, sec.\ 65 \& 91]{becker1982}).  Morozov's solution to the $4/3$ problem is somewhat similar to the one presented here, though he does not use the concept of field velocity.}  The energy in the electromagnetic field does not merely need to keep up with the motion of the sphere.  It also must travel from the back of the sphere to the front (as shown in figures \ref{V2fig} and \ref{G2fig}).  Thus, the field's average velocity must exceed the velocity $\vec{v}$ with which the sphere moves, resolving the revised puzzle raised at the end of the previous section.

\newpage

To understand quantitatively why the electromagnetic field energy must move at $\frac{4}{3}$ the velocity of the sphere, let us analyze the local conservation of energy within the sphere:
\begin{align}
\frac{\partial \rho_s^{\mathcal{E}}}{\partial t}&= -c^2\vec{\nabla}\cdot\vec{G}_s+\vec{E}\cdot\vec{J}
\ ,
\end{align}
where $\rho_s^{\mathcal{E}}$ is the sphere's energy density, $\vec{G}_s$ is the sphere's momentum density, and $\vec{G}_s c^2$ is the sphere's energy flux density.  We can separate the momentum density $\vec{G}_s$ into a part 1, which describes the flow of energy forward at constant velocity $\vec{v}$ as the sphere moves along, and a part 2, which describes the backward flow of energy that is needed to move the energy gained from the electromagnetic field in the front half of the sphere to the back half of the sphere where it is emitted,
\begin{align}
\vec{G}_{s}&=\vec{G}_{s1}+\vec{G}_{s2}
\nonumber
\\
\vec{\nabla}\cdot\vec{G}_{s1}&=\frac{-1}{c^2}\frac{\partial \rho_s^{\mathcal{E}}}{\partial t}
\nonumber
\\
\vec{\nabla}\cdot\vec{G}_{s2}&=\frac{1}{c^2}\vec{E}\cdot\vec{J}
\ .
\label{momentumdecomposition}
\end{align}
We can calculate the total momentum associated with the backward flow of energy by integrating $\vec{G}_{s2}$ using integration by parts (where the surface term vanished because $\vec{G}_{s2}$ is zero outside the sphere), \eqref{EdotJ}, and \eqref{momentumdecomposition},
\begin{align}
\int d^3\vec{x} \ \vec{G}_{s2}&=\int d^3\vec{x} \ (\vec{G}_{s2}\cdot\vec{\nabla}) \vec{x}
\nonumber
\\
&=- \int d^3\vec{x} \  (\vec{\nabla}\cdot \vec{G}_{s2}) \vec{x}
\nonumber
\\
&=\frac{-1}{c^2} \int d^3\vec{x} \ (\vec{E}\cdot\vec{J}\,) \vec{x}
\nonumber
\\
&=-\int_0^R dr \int_0^\pi d\theta \ \frac{3 Q^2 r^4}{2R^6 c^2}\cos^2(\theta)\sin(\theta) \vec{v}
\nonumber
\\
&=-\frac{Q^2}{5Rc^2}\vec{v}
\ .
\label{Gs2}
\end{align}
Only if the average velocity of the field is $\frac{4}{3}\vec{v}$ will the flow of energy in the electromagnetic field, with momentum $m_{em}\langle \vec{v}_f \rangle$, be calibrated to balance this flow of energy in matter so that
\begin{align}
m_{em}\langle \vec{v}_f \rangle + \int d^3\vec{x} \ \vec{G}_{s2} &= m_{em}\vec{v}
\nonumber
\\
\frac{3Q^2}{5 R c^2}\langle \vec{v}_f \rangle -\frac{Q^2}{5Rc^2}\vec{v} &= \frac{3Q^2}{5 R c^2}\vec{v}
\label{theargument}
\end{align}
and the total momentum of the system, $m_{em}\langle \vec{v}_f \rangle +\int d^3\vec{x} \ \vec{G}_{s}$, is $m\vec{v}=(m_{em}+m_s)\vec{v}$.

In the above story, $\vec{v}$ is the velocity at which the sphere's charge moves and $\vec{G}_{s1}=m_s\vec{v}$.  One could also introduce a velocity of energy flow for the sphere (analogous to the velocity of energy flow in the electromagnetic field).  The average velocity of energy flow would be $\vec{G}_{s}/m_s$, which would be less than $\vec{v}$ because of the backwards flow of energy in $\vec{G}_{s2}$.

In section \ref{standardpuzzlesection}, the standard puzzle was presented as a contradiction between two ways of finding the electromagnetic mass of the field surrounding a charged body.  Some authors discuss a third way of finding the electromagnetic mass, by examining the force that is required to accelerate the charged body \cite{griffithsowen1983}.  This self-force-derived mass is found to be $\frac{4}{3}m_{em}$ from the assumption that the electromagnetic field is being accelerated to the same velocity as the body.  Recognizing that enough force must be imparted to accelerate the electromagnetic field to $\frac{4}{3}\vec{v}$ clarifies that the correctly self-force-derived mass is $m_{em}$.

Before proceeding, let us take a moment to briefly consider the role of the cohesive forces (or ``Poincar\'{e} stresses'') holding the charged sphere together, by asking what would happen if they were absent.  Consider a sphere of uniform charge density with no cohesive forces, initially moving at constant velocity $\vec{v}$ surrounded by the electromagnetic field in \eqref{Efieldsphere} and \eqref{Bfieldsphere} (with average field velocity of $\frac{4}{3}\vec{v}$).  As the sphere moves forward, the forces of electric self-repulsion would cause it to rapidly expand (converting electromagnetic field energy into kinetic energy as the electromagnetic field weakens).  The front half of the moving expanding sphere would gain more kinetic energy than the front half of an otherwise similar expanding stationary sphere because the expansion and the forward motion are aligned (and kinetic energy is proportional to velocity squared).  The back half would gain less.  Thus, we see that there is still a need for the energy in the electromagnetic field energy to flow around from back to front as the sphere expands.  However, there is no need for energy to be transferred within the expanding sphere from front to back.  It is only when we have cohesive forces opposing the expansion that we must have a flow of energy within the sphere opposite the direction of motion.

\section{Conclusion}\label{conclusionsection}

We have seen that the $4/3$ problem for mass renormalization in classical electromagnetism can be resolved by noting that the electromagnetic field's velocity is faster than the charged body's, and that there must be a flow of energy within the body from its front towards its back (for the body to retain its shape while respecting local conservation of energy).

One reason for studying mass renormalization in classical electromagnetism is to understand what it might tell us about mass renormalization in a quantum field theory like quantum electrodynamics.  As a step towards understanding mass renormalization for the electron in quantum electrodynamics, one might consider a classical theory of interacting electromagnetic and Dirac fields (Maxwell-Dirac theory) where energy and momentum are both locally conserved.  The Dirac field possesses densities of energy and charge, as well as velocities of energy and charge flow \cite{howelectronsspin, smallelectronstates, bb2021, fields2022}.  In this context, one could improve on the classical model of the electron as a charged sphere (studied here) and analyze a Dirac field wave packet moving forward at a low velocity (and interacting with its own electromagnetic field) to see whether there is any backwards flow of energy in the wave packet (as occurs in the charged sphere) or none (as occurs when you remove the cohesive forces and allow the sphere to freely expand under its own self-repulsion, as was discussed at the end of section \ref{resolutionsection}).  Moving from this classical field theory to quantum electrodynamics, the story changes.  The forces of electric self-repulsion, which were central to the account of energy flow in section \ref{resolutionsection}, appear to be eliminated \cite{selfrepulsion}.

\textbf{Acknowledgments}
Thank you to Alex Blum and Mario Hubert for helpful feedback.

\end{document}